\documentstyle[aps,prl,multicol,epsfig]{revtex}

\newcommand{\be}{\begin{eqnarray}}

\newcommand{\ee}{\end{eqnarray}}
\def\ket#1{|#1\rangle}

\draft
\tighten

\begin{document}
\title{Laser probing of Cooper-paired trapped atoms}

\author{G.M.\ Bruun$^1$, P.\ T\"orm\"a$^2$, M.\ Rodriguez$^2$ and P.\ Zoller$^3$}
\address{$^1$Nordita, Blegdamsvej 17, 2100 Copenhagen, Denmark\\
$^2$Laboratory of Computational Engineering, P.O.Box 9400, FIN-02015
Helsinki University of Technology, Finland\\
$^3$Institute for Theoretical Physics, University of Innsbruck,
Technikerstra{\ss}e 25, A-6020 Innsbruck, Austria}
\maketitle

\begin{abstract} 
We consider a gas of trapped Cooper-paired fermionic atoms 
which are manipulated by laser light. The laser induces a transition
from an internal state with large negative scattering length (superfluid) 
to one with weaker interactions (normal gas). We show that the process can be used 
to detect the presence of the superconducting order parameter. Also, we propose a direct way of 
measuring the size of the gap in the trap. The efficiency
and feasibility of this probing method is investigated in detail in different physical situations.
\end{abstract}


\begin{multicols}{2}[]

\section{Introduction}

Recent experiments on cooling and trapping Fermionic atoms
have opened up new opportunities for
studying fundamental quantum statistical and many-body physics.
Trapped Fermionic $^{40}$K atoms were cooled down to
temperatures where the Fermi degeneracy sets in
\cite{Jin99}. Two lithium isotopes were trapped simultaneously
in an magneto-optical trap in \cite{Mewes00} and optical trapping of 
fermionic lithium has been achieved as well \cite{Thomas00}. 
The richness of the internal energy structure of the atoms and the
possibility to accurately and efficiently manipulate these energy
states by laser light and magnetic fields allows excellent control
of these gases. Furthermore, atomic gases
are dilute and weakly interacting thus offering the ideal tool for
developing and experimentally testing theories of many-body quantum physics.

The degenerate Fermi gas is expected to show many interesting phenomena 
in its thermodynamics\cite{BruunBurnett}, excitation spectrum 
\cite{QHE,BruunClark,Stringari,Csordas}, collisional dynamics\cite{Coll} and
scattering of light\cite{Jin99,Janne99,Zoller}.
A major goal is to observe the predicted \cite{Stoof,You99} 
BCS-transition for Fermionic atoms -- this would compare to the experimental
realization of atomic Bose-Einstein condensates \cite{Bose}.  It is
still, however, an open question how to observe the BCS-transition,
because the value of the superconducting order parameter (gap) is
expected to be small and the existence of the order parameter does
not significantly change the density profile and other bulk properties 
of the gas.  

There are several proposals for measuring the superconducting order
parameter. Off-resonant light scattering as a probe was proposed
in \cite{Weig99,Zhang99}. Superfluidity is predicted to effect both
the spectral and spatial distribution of the scattered light. 
In \cite{Ruostekoski99}, superfluidity was found to increase
the optical lineshift and linewidth.
Also non-optical phenomena, such as collective
and single particle excitations, have been proposed to be used for
observing the BCS-transition 
\cite{Baranov98,Baranov99,Bruun99,Zambelli00,Minguzzi00}.
Probing by a magnetic field was considered in \cite{Petrosyan99}.

The use of on-resonant light as a probe for the order parameter was
proposed in \cite{PTPZ}. The basic idea is to transfer atoms from one
internal (hyperfine) state for which the atoms are Cooper paired to
another state for which the interatomic interaction is not strong
enough to lead to a BCS state. This effectively creates a
superconducting -- normal state interface across which the atomic
population can move. There is a conceptual analogy to electron
tunneling from a superconducting metal to a normal one which is used
to measure the gap and the density of states for Cooper-paired
electrons \cite{Mahan}. The tuneability of the interaction strengths which is
required for this scheme is obtained by the use of magnetic fields.
This allows to manipulate the scattering lengths between atoms in
different internal states, see e.g.\ the recent experimental results
concerning optically trapped fermionic Lithium atoms \cite{Thomas00}
and the theory predictions for $^{40}K$ \cite{Bohn00}.

The basic idea in the proposal \cite{PTPZ} is that the absorption peak
is shifted and becomes asymmetric because of the existence of the gap
-- the laser has to provide energy for breaking the Cooper pairs in
order to transfer atoms from the paired state to the unpaired one.
This behaviour is, however, strongly influenced by the specific
physical situation. In this paper we investigate in detail how the
choice of the chemical potentials for the superfluid and the normal
state, and the choice of the interaction strengths and laser profiles
affect the absorption. We also compare the results in the cases of a
homogeneous system and a trapped gas. In section \ref{SNinterface}, we
introduce the considered system. The linear response of the gas for a
light probe is derived in section \ref{LinResp}, both for a
homogeneous and a trapped gas. In section \ref{Effect} the various
parameters that affect the observed absorption are discussed. In
section \ref{Const} exact numerical results are presented for the
limit when the laser beam profile can be considered a constant.  
A beam profile with non-zero intensity only in a small volume in the
middle of the trap is considered in section \ref{Sphere}. 
This case is very interesting because then only the
center of the trap is probed. Thus the order parameter seen by the
laser is almost constant and indeed, we find a remarkable agreement
with the results predicted for a homogeneous system. We finally
summarize the results in section \ref{Conclusions}.

\section{The superconductor - normal state interface}\label{SNinterface}

\narrowtext
\vbox{
\begin{figure}
\begin{center}
\epsfig{file=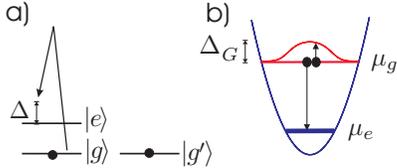,width=2.0in,angle=0} 
\end{center}
\caption{Probing of the gap in a gas of attractively interacting cold
Fermionic atoms. a) Laser excitation with the coupling $\Omega$ and
the detuning $\Delta$ transfers a Cooper paired atom from
the internal state $\ket{g}$ to the state $\ket{e}$.  b) The other
atom in the initial Cooper pair becomes an excitation in the BCS
state, therefore the laser has to provide also the additional gap
energy $\Delta_G$.  In this picture the Fermi levels $\mu_g$ and 
$\mu_e$ for the internal states have been chosen to be different from each other but
they could also equal.} 
\end{figure} }

We consider atoms with three internal states available, say $\ket{e}$,
$\ket{g}$, and $\ket{g'}$. They are chosen so that the interaction
between atoms in states $\ket{g}$ and $\ket{g'}$ is relatively strong
and their chemical potentials are nearly equal so that $\ket{g}$ and $\ket{g'}$
can be assumed to be Cooper-paired. All other interactions are small
enough and/or the chemical potentials of the corresponding states are 
different enough in order to assume the $\ket{e}$ atoms
to be in a normal state \cite{Stoof}. The laser
frequency is chosen to transfer population between $\ket{e}$ and
$\ket{g}$, but is not in resonance with any transition which could
move population away from the state $\ket{g'}$. 

For small intensities, the laser-interaction can be
treated as a perturbation, the unperturbed states being the normal and
the superconducting state. The transfer of atoms from $\ket{g}$ to
$\ket{e}$ is then analogous to tunneling of electrons from a normal
metal to a superconductor induced by an external voltage, which can be
used as a method to probe the gap and the density of states of the
superconductor \cite{Mahan}. In our case the tunneling is between
two internal states rather than two spatial regions, this
resembles the idea of internal Josephson-oscillations in
two-component Bose-Einstein condensates \cite{Williams99}.
Fig.\  1 illustrates the basic idea.  
The observable carrying essential information about the superconducting
state is the change in the population of the state $\ket{e}$, we call
this the current $I$.  

We define a two-component Fermion field

\be
\psi ({\bf x}) = 
\left(\begin{array}{c} \psi_e({\bf x}) \\ \psi_g({\bf x})
\end{array} \right) ,
\ee
where $\psi_e$ and $\psi_g$ fulfill standard Fermionic
commutation relations. The fields $\psi_{e/g}$ can be expanded
using some basis functions (e.g. plane waves or trap wave functions) and 
corresponding creation and annihilation operators:
$
\psi_{e/g}({\bf x})= \sum_j c_j^{e/g} \phi_j^{e/g} 
({\bf x})
$.
The annihilation and creation operators fulfil $\{ {c_i^e}^\dagger,
c_j^g\}=0$ and $\{ c_i^{e/g\dagger},c_j^{e/g}\}=\delta_{ij}$. 
The two components of the field, corresponding to the internal
states $\ket{e}$ and $\ket{g}$, are coupled by a laser.
This can be either a direct excitation or a Raman process;
we denote the atomic energy level difference 
by $\omega_a$ ($\hbar\equiv1$), the laser frequency 
$\omega_L$ and the wave vector
$k_L$ -- in the case of a Raman process these are effective
quantities. In the rotating wave approximation the Hamiltonian reads
\be
H &=& H_e + H_{gg'} + \frac{\Delta}{2} \int d^3x \psi^\dagger ({\bf x}) \sigma_z
\psi ({\bf x}) \nonumber \\
&&+  \int d^3x \psi^\dagger ({\bf x})
\left( \begin{array}{cc}
0 & \Omega({\bf x}) \\
\Omega^*({\bf x}) & 0 
\end{array} \right) 
\psi ({\bf x}) .
\ee
Here $\Delta = \omega_a - \omega_L$ is the (effective) detuning, and
$\Omega({\bf x})$ contains the spatial dependence of the laser field
multiplied with the (effective) Rabi frequency.  The parts $H_e$ and
$H_{gg'}$ contain terms which depend only on $\psi_e$ or $\psi_g$, $\psi_{g'}$,
respectively. Possible spatial inhomogeneity,
e.g.\ from the trap potential, is also included in $H_e$ and
$H_{gg'}$.

\section{The current}\label{LinResp}

The observable carrying essential information about
the superconducting state is the rate of change in 
population of the $\ket{e}$ state. 
We may call it, after the electron tunneling analogy, 
the current $$I(t) = - \langle \dot{N_e} \rangle,$$ where
$
N_e = \int d^3x  \psi_e^\dagger ({\bf x}) \psi_e ({\bf x})
$.
The current $I(t)$ is calculated considering the tunneling part of
the Hamiltonian, $$H_T = H - (H_e + H_{gg'} + \Delta/2 (N_e - N_g))$$
as a perturbation;
the current $I$ becomes the first order response 
to the external perturbation caused by the laser. We calculate
it both in the homogeneous case and in the case of harmonic confinement.
The calculations are done in the grand canonical ensemble, therefore 
the chemical potentials $\mu_g$ and $\mu_e$ are introduced. 
Also the detuning $\Delta$ acts like
a difference in chemical potentials, thus it becomes useful to 
define an effective quantity of the form $\tilde{\Delta} = \mu_e
- \mu_g + \Delta \equiv \Delta \mu + \Delta$. In the derivation
we assume finite temperature, but most of the results will only 
be quoted for T=0.

\subsection{Homogeneous case} \label{Homogeneous}

The assumption of spatial homogeneity is
appropriate when the atoms are confined in a trap
potential which changes very little compared to characteristic
quantities of the system, such as the coherence length and 
the size of the Cooper pairs. In the present context, this assumption
is also valid when the laser profile is chosen so that it only probes the
middle of the trap where the order parameter is nearly 
constant in space -- this will be discussed in more detail
in section \ref{Sphere}.

In the homogeneous case the Fermion fields $\psi_{e/g}$
can be expanded into plane waves. The Hamiltonian becomes
\be
H &=& H_e + H_{gg'} + \frac{\Delta}{2} \sum_k [{c_k^e}^\dagger c_k^e
- {c_k^g}^\dagger c_k^g] \nonumber \\
&&+ \sum_{kl} [T_{kl} {c_k^e}^\dagger c_l^g
+ h.c.] ,  \label{ham2}
\ee
where
$$
T_{kl} = \frac{1}{V} \int d^3x 
\Omega ({\bf x}) e^{i{\bf k} \cdot {\bf x}} e^{-i{\bf l} \cdot {\bf x}}.
$$
We calculate the current 
\be
I=-\langle \dot{N_e} \rangle = -i \langle 
[H,N_e] \rangle
\ee
treating $H_T$ as a perturbation: terms of higher
order than $H_T^2$ are neglected. Because we are interested in the
current between the superconducting and normal states, correlations
of the form $\langle c_e^\dagger c_e^\dagger c_g c_g \rangle$ 
(and $h.c.$) are omitted since they correspond to tunneling of 
pairs (Josephson current). The current can be written
\be
I = \int_{-\infty}^\infty dt \theta (t) 
(e^{-i\tilde{\Delta}t} \langle [A^\dagger(0),A(t)] \rangle
-e^{i\tilde{\Delta}t} \langle
[A(0),A^\dagger(t)] \rangle ) \nonumber
\ee
where $A(t) = \sum_{kl} T_{kl} {c_k^g}^\dagger(t) c_l^e(t)$, and
$c_l^{e/g}(t) = e^{iKt} c_l^{e/g} e^{-iKt}$ where $K=H-H_T-\mu_e N_e
-\mu_g N_g$. The two terms in the above equation have the form of
retarded and advanced Green's functions. These are evaluated
using Matsubara Green's functions techniques, which leads to 
$$
I = \sum_{kl} |T_{kl}|^2 \int_{-\infty}^\infty \frac{d\epsilon}{2\pi}
[n_F(\epsilon) - n_F(\epsilon + \tilde{\Delta})]
A_g(k,\epsilon + \tilde{\Delta})A_e(l,\epsilon) , 
$$
where $n_F$ are the Fermi distribution functions and $A_{g/e}$ 
are the spectral functions for the superconducting and normal
states. We use the standard expressions \cite{Mahan}
$$A_{e}(l, \epsilon) = 
2\pi \delta (\epsilon - \xi_l)$$ and $$A_{g}(k, \epsilon +
\tilde{\Delta}) = 2 \pi [u_k^2 \delta (\epsilon + \tilde{\Delta}
- \omega_k) + v_k^2 \delta (\epsilon + \tilde{\Delta} + \omega_k)].$$
Here $\xi_l = E_l - \mu_e$ and $u_k$, $v_k$ and 
$\omega_k$ are given by the Bogoliubov transformation.
Here we consider for simplicity the term proportional to $v_k^2$;
$u_k^2$ is analogous. The laser field is chosen to be a running wave, that
is $\Omega ({\bf x}) = \Omega e^{i {\bf k_L} \cdot {\bf x}}$. The
term $|T_{kl}|^2$ now produces a delta-function enforcing momentum
conservation. Note that this is very different from the 
assumption of a constant transfer matrix ($\sum_{kl} |T_{kl}|^2 
\longrightarrow |T|^2 \sum_{kl}$) made in the standard calculation
for tunneling of electrons over a superconductor -- normal metal
surface \cite{Mahan}. The final result becomes (assuming for simplicity
that the temperature $T=0$)
\be
I = - \pi \Omega^2 \rho(\Delta) \theta (-\tilde{\Delta}-\omega_{\tilde{k}-k_L} -
\Delta \mu ) \frac{\omega_{\tilde{k}-k_L} -
\xi_{\tilde{k}-k_L}}{\omega_{\tilde{k}-k_L} 
+ \xi_{\tilde{k}-k_L}\left[1-\frac{k_L}{\tilde{k}}\right]}   \nonumber
\ee
where $\tilde{k}$ is given by the following energy conservation
condition:
$$-\tilde{\Delta}+\omega_{\tilde{k}-k_L} + \xi_{\tilde{k}-k_L} = 0,$$
$\omega_k = \sqrt{\xi_k^2 + \Delta_G^2}$, and 
$$\rho(\Delta) = \frac{V}{2\pi^2}\sqrt{\frac{\Delta_G^2-\Delta^2}
{\Delta}+2\mu_g}$$ 
is the density of states which appears when the summation over momenta is
changed into an integration over energies.

The laser momentum $k_L$ can be very small compared to the momentum
of the atoms, especially in the case of a Raman process. By setting
$k_L=0$ the result becomes, including now terms proportional to 
both $v_k^2$ and $u_k^2$: 
\be
I = \pm \pi \Omega^2 \rho(\Delta) \theta (\Delta^2 - \Delta_G^2 + 2
\Delta \mu \Delta) \frac{\Delta_G^2}{\Delta^2}  ,
\label{peak}
\ee
where $\pm$ are for $\Delta > 0$ and $\Delta < 0$, respectively.
The term with the $-$ sign corresponds to ${\Delta}<0$ i.e.\ current
from $\ket{g}$ to $\ket{e}$, and the positive term to current
from $\ket{e}$ to $\ket{g}$. 
To understand the results in terms of physics, let us first consider
the case of equal chemical potentials $\Delta \mu = 0$:
\be
I =  -  \pi \Omega^2 \frac{\Delta_G^2}{\Delta^2}\rho(\Delta)[\theta (-\Delta - \Delta_G)
-\theta (\Delta - \Delta_G)]
.  \label{curr2}
\ee
In order to transfer one atom from the state $\ket{g}$ to $\ket{e}$
the laser has to break a Cooper pair. The minimum energy required for this
is the gap energy $\Delta_G$, therefore the current does not flow 
before the laser detuning provides this energy -- this is expressed
by the first step function in (\ref{curr2}). As $|\Delta|$ increases
further, the current will decrease quadratically. This is because 
the case $|\Delta|=\Delta_G$ corresponds to the transfer of particles
with $p=p_F$, whereas larger $|\Delta|$ means larger momenta, and there
are simply fewer Cooper-pairs away from the Fermi surface. This
behaviour is very different from the electron tunneling where the current grows
as $\sqrt{(eV)^2 - \Delta_G^2}$ \cite{Mahan} (the voltage $eV$ corresponds to
the detuning $\Delta$ in our case) because all momentum states are coupled
to each other.  The second step function in (\ref{curr2}) corresponds to tunneling into the
superconductor. In this case one has to provide extra energy because a single
particle tunneling into a superconductor becomes a quasi-particle excitation
with the minimum energy given by the gap energy. 

When the chemical potentials are not equal the situation is more
complicated, but the basic features are the same: i) threshold for the
onset of the current determined by the gap energy and difference in chemical
 potentials, and ii) further decay
of the current because the density of the states that can fulfill energy
and momentum conservation decreases.

\subsection{Harmonic confinement}

In the case of harmonic confinement the spatial dependence
of the current is non-trivial. We define the total current
as
\be
I(t) = - \int d^3x \langle \dot{N_e} ({\bf x}) \rangle ,
\ee
where
\be
\dot{N_e} ({\bf x}) = i [H,N_e({\bf x})] = 
i [\Omega^*({\bf x}) \psi^\dagger_g ({\bf x}) \psi_e ({\bf x}) - H.c.] .
\ee
No expansion in the plane waves or other basis functions is made
at this point and the first order perturbation calculation leads
to a result with explicitly spatially dependent correlation
functions:
\be
I = 2 Im
[X_{ret}(-\tilde{\Delta})] , 
\ee
where
\be
X_{ret}(-\tilde{\Delta})& =& i \int_{-\infty}^\infty dt e^{-i\tilde{\Delta}t}
\theta (t) \int d^3x \nonumber \\&& \times \int d^3x' \langle [A^\dagger ({\bf x}, 0),
A({\bf x'},t)] \rangle ,
\ee
and
\be
A({\bf x},t) = \Omega^*({\bf x}) \psi^\dagger_g ({\bf x}) \psi_e ({\bf x}) .
\ee
The retarded Green's function $X_{ret}(-\tilde{\Delta})$ is calculated
using Matsubara techniques. For performing the Matsubara summations,
the spatially dependent Green's functions for the normal and the superfluid
state are expanded using the trap wave functions $\phi_n ({\bf x})$, and
the Bogoliubov-de Gennes (BdG) eigenfunctions $u_n({\bf x})$ and $v_n({\bf x})$,
respectively. This leads to the result
\be
&&X_{ret}(-\tilde{\Delta}) = \int_{-\infty}^\infty
\frac{d\epsilon}{2\pi} \int d^3 x \int d^3 x' \Omega^*({\bf x})
\Omega({\bf x'}) \nonumber \\ && [\tilde{A}_e({\bf x}, {\bf x'}, \epsilon) 
G_{adv}^g({\bf x'}, {\bf x}, \epsilon + \tilde{\Delta}) \nonumber \\ && 
+ G_{ret}^e({\bf x}, {\bf x'}, \epsilon - \tilde{\Delta})
\tilde{A}_g({\bf x'}, {\bf x}, \epsilon)] ,  \label{curr3}
\ee
where $\tilde{A}_{e/g}$ are defined
$$\tilde{A}_{e/g}({\bf x}, {\bf x'}, \epsilon ) = i (G_{ret}^{e/g}({\bf
x}, {\bf x'}, \epsilon)
- G_{adv}^{e/g}({\bf x}, {\bf x'}, \epsilon)) ,$$
and
\be
G_{adv}^{g}({\bf x'}, {\bf x}, \epsilon) &=& \sum_n \frac{u_n({\bf
x'})u_n^*({\bf x})}{\epsilon - \omega_n - i \delta}
+ \frac{v_n^*({\bf
x'})v_n({\bf x})}{\epsilon + \omega_n - i \delta}  \\
G_{ret}^{e}({\bf x}, {\bf x'}, \epsilon) &=& \sum_n \frac{\phi_n ({\bf x})
\phi_n^* ({\bf x'})}{\epsilon - \xi_n + i \delta} .
\ee
As mentioned, $u_n({\bf x})$, $v_n({\bf x})$ and $\omega_n$ are given by the 
BdG equations \cite{deGennes}.
In taking the imaginary part of the expression (\ref{curr3}) we first
collect together all spatially dependent terms, which gives real factors
of the form $|\int d^3 x \Omega ({\bf x}) u_n ({\bf x})\phi_m({\bf x})|^2$
(we also use the fact that the trap wave functions $\phi_m({\bf x})$ are real).
Imaginary parts of the remaining terms give spectral functions in the
usual way. The derivation leads to
\be
I &=& - 2\pi \sum_{n,m} \left|\int d^3 x \Omega ({\bf x}) u_n ({\bf x})
\phi_m({\bf x})\right|^2 [n_F(\omega_n) - n_F(\xi_m)] \nonumber \\ &&\delta (\xi_m +
\tilde{\Delta} - \omega_n) + \left|\int d^3 x \Omega ({\bf x}) v_n^* ({\bf x})
\phi_m({\bf x})\right|^2 \nonumber \\ && [n_F(-\omega_n) - n_F(\xi_m)] \delta (\xi_m +
\tilde{\Delta} + \omega_n)  .  \label{final}
\ee
Note that this form does not lead to a simple step-function type
behaviour like in the homogeneous case. Due to the non-orthogonality 
of the trap and the BdG wavefunctions transitions
between many quantum numbers are allowed and the total current is
the sum of all these. 

In the following sections we use the result (\ref{final}) to investigate the
feasibility of the method as a probe in different physical situations.
The eigenfunctions and values $v_n({\bf x})$ and $\omega_n$ are 
calculated numerically from the BdG equations using
the pseudo-potential method presented in \cite{Georg}. We 
assume for simplicity  that the trap has spherical symmetry. The 
quantum numbers $n,m$ in Eq.(\ref{final}) then become $\eta,l,m$ with 
$l,m$ being the usual angular momentum quantum numbers. The quasi particle (QP)
energies will only depend on $\eta,l$. The method of solving the BdG equations is
described in detail in \cite{Georg}.

Note that the above derivation assumes that the BdG
equations are solved exactly. One can, however, also use the local density
approximation \cite{Stoof97} where the chemical potential of the superconducting
state is assumed to have a parametric dependence on position. In this
case one should repeat the above derivation by assuming that
the superfluid state Hamiltonian $H_{gg'}$ is parametrically depended
on ${\bf x}$: $H_{gg'}^{\bf x}$. This leads to a result identical to 
(\ref{final}) except that now $u_n$, $v_n$ are actually
plane waves but they as well as $\omega_n$ have a parametric dependence
on ${\bf x}$ via $\mu ({\bf x})$. We may denote the current for a chosen ${\bf x}$ as
$I^{\bf x}$. The total current is then the average of $I^{\bf x}$ for 
all ${\bf x}$.

\section{The effect of chemical potential, laser profile and trapping}\label{Effect}

The behaviour of the current is strongly influenced by the
choice of the physical parameters. This allows a convenient
way to optimize the probing scheme as well as to investigate 
interesting physics such as the influence of the harmonic confinement
--- this is our twofold aim in the following.

Although for instance the laser profile can be chosen by will,
there are a few basic restrictions in choosing the chemical potentials
and the interaction strengths between the atoms in different internal states.
In order for a gas of Fermionic atoms in two internal states to form
Cooper pairs, the interatomic interaction should be large enough
and the chemical potentials corresponding to the two states should
be very close to each other \cite{Stoof}. This is the condition we
assume for $\ket{g}$ and $\ket{g'}$. The state $\ket{e}$ is always
assumed to have either negligible interaction with the states  
$\ket{g}$ and $\ket{g'}$, or considerably lower chemical potential
(smaller number of particles). It turns out that not only the pairing 
effects the light absorption but also normal interactions described by
the Hartree-field are crucial. This is illuminated by a comparison
of these two cases of $\ket{e}$ having negligible scattering length
or small chemical potential. 

In real experiments, the gas is trapped in a harmonic potential.
We have derived results also for the homogeneous case, these can
be used in the case of very large traps and they also give a simple
intuitive picture of the basic physics in this system. We will show
that indeed the trapping has considerable effect on the results.
On the other hand, an effectively homogeneous situation can
be achieved by choosing the laser to probe only the center of the
trap. As will be shown, this avoids certain problems arising from
the non-homogeneous potential and can give a very clear signature
of the superconducting state.

\section{Constant Beam Profile}\label{Const}
In this section we present exact numerical results in  the limit when the laser beam intensity 
can be well approximated by a constant. That is, we take
  $\Omega({\mathbf{x}})=\Omega$ in Eq.(\ref{final}).
 As the typical effective wavelength for the laser is much larger than the extend of
the trap for the relevant energies, such an approximation should be 
good as long as the laser amplitude is constant over the whole extend of the cloud.
We present results for 
two situations: the case when there are no atoms initially in the state $\ket{e}$ and the
case when the chemical potentials for $\ket{e}$ and $\ket{g}$ are the same.

\subsection{No $\ket{e}$ atoms initially}\label{mue0}
\begin{figure}
\centering
\epsfig{file=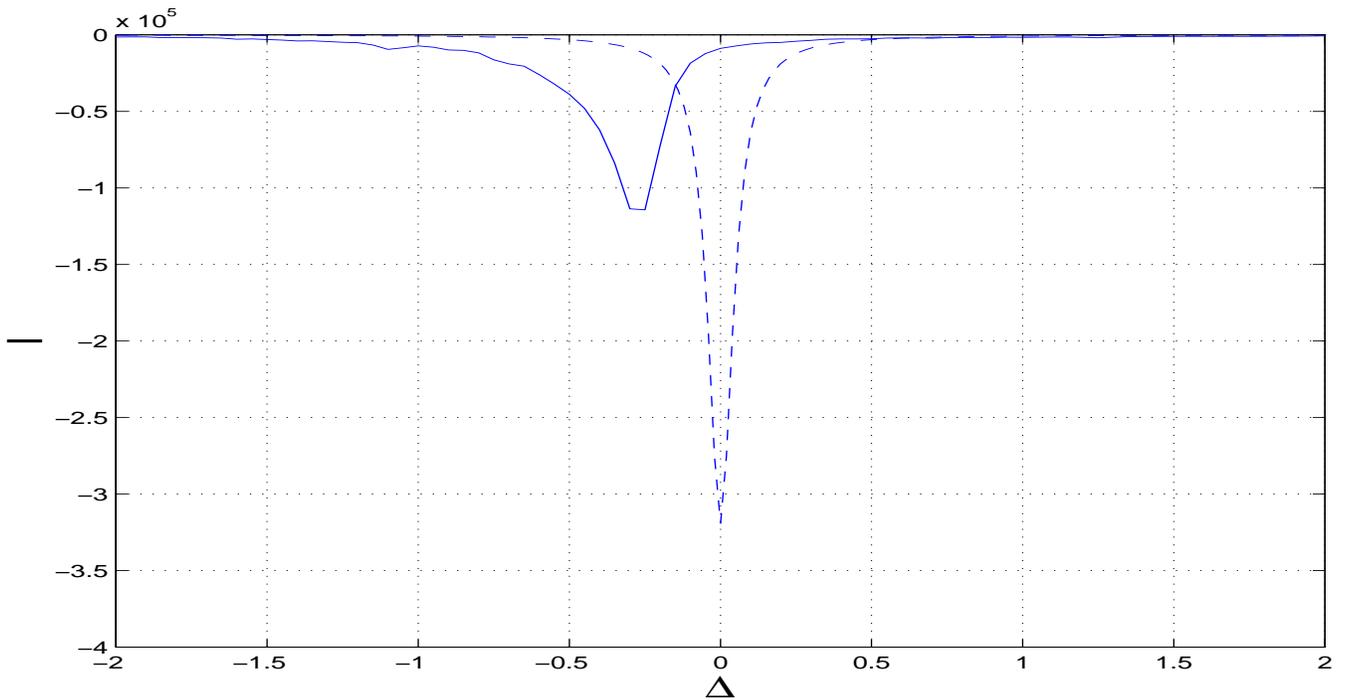,height=0.4\textheight,width=\textwidth}
\caption{The current $I=-\langle\dot{N}_e\rangle$ as a function of the laser detuning for 
$\mu_g=31.5$, $\mu_e=0$ (in units of 
 $\omega$), and  $g_{eg}+g_{eg'}=g_{gg'}$.}
\label{mug315eempty}

\end{figure}

In the case considered in this subsection, 
we assume to have initially only a gas of interacting $\ket{g}$ and 
$\ket{g'}$ atoms. The laser beam then induces transitions to the 
hyperfine level $\ket{e}$. To obtain clear results, we will assume that 
the $\ket{e}$ atoms
see the same Hartree field as the $\ket{g}$ atoms. This means that
$g_{eg}+g_{eg'}=g_{gg'}$ since the $\ket{e}$ atoms see the 
Hartree field from both the $\ket{g}$ and
$\ket{g'}$ atoms. Here, $g_{gg'}=4\pi a_{gg'}/m$ denotes the interaction 
strength between the two hyperfine states $\ket{g}$ and $\ket{g'}$
and likewise for $g_{eg}$ and $g_{eg'}$. The parameter 
$a_{gg'}$ is the usual $s-$wave scattering 
length for scattering between $\ket{g}$ and $\ket{g'}$ atoms.
Experimentally, this situation could possibly be achieved by manipulating an external 
magnetic field  thereby tuning the effective low energy interaction between the relevant
hyperfine states to an appropriate value. In Fig.\ \ref{mug315eempty}, we show 
a typical example of the current $I(\Delta)=-\langle\dot{N_e}\rangle$.
 We have used parameters such that 
$g_{gg'}=-l_{ho}^3\omega$, $\mu_g=31.5\omega$ and the temperature $T=0$ with  $\mu_g$
denoting the chemical potential for the $\ket{g}$ and $\ket{g'}$ atoms. 
Here $l_{ho}=(m\omega)^{-1/2}$ is the harmonic oscillator length. For $^6$Li atoms with 
$a_{gg'}$=-1140\AA~\cite{Abraham}, this corresponds to approx.\ $1.6\times10^4$ atoms trapped
 in the states $\ket{g}$ and $\ket{g'}$ with a trapping frequency of 820Hz yielding a critical 
temperature  $T_c\sim 110nK$ for the BCS transition.  
We have added an imaginary part $\Gamma=0.1\hbar\omega$ to the quasiparticle energies 
such that the $\pi\delta(x)$-functions in Eq.(\ref{final}) become Lorentzians 
$\Gamma/2(x^2+\Gamma^2/4)$.

{\it Normal-normal current}
The dashed curve in Fig.\ \ref{mug315eempty} depicts the current when the
$\ket{g}$ and $\ket{g'}$ atoms are in the normal phase. Since the 
$\ket{e} $ atoms see the same Hartree-field as the $\ket{g}$ 
atoms, the spatial part of the QP wavefunctions is the same 
for the two hyperfine states. Thus, the overlap integrals in Eq.(\ref{final}) simply 
become $\delta_{nm}$-functions and since the quasiparticle spectrum is 
the same for the two hyperfine states, we obtain from Eq.(\ref{final}):
\begin{equation}\label{Lorentz}
I(\Delta)=-2N_g\frac{\Gamma/2}{\Delta^2+\Gamma^2/4}.
\end{equation} 
where $N_g$ is the number of $\ket{g}$ atoms trapped. 

{\it Superconductor-normal current}
The solid curve Fig.\ \ref{mug315eempty} depicts the current when $\ket{g}$ and $\ket{g'}$ are
in the superfluid phase. We see that the maximum of the current is displaced
from $\Delta=0$ and that the shape of the current profile is asymmetric.
Both effects are quite straightforward to understand. The asymmetry reflects 
the fact that the current now is given by a sum of Lorentzians centered at 
different frequencies since the QP spectra for $\ket{g}$ and $\ket{e}$
are different and the overlap integral in Eq.(\ref{final}) does not give a simple 
selection rule. The shift in the center of the peak to negative $\Delta$
is due to the fact that in order to induce a transition from a low lying 
QP-state in the superfluid one needs to break a Cooper pair. This requires 
an additional energy given by the pairing energy of the QP-state.
As a fraction $\sim T_c/T_F$  of the particles participates in the
 pairing and  they on average have the pairing energy $T_c$, one 
can estimate the order of magnitude of the  
shift in the center of the peak away from its normal phase value
$\Delta=0$ to be ${\mathcal{O}}(T_c^2/T_F)$.
 Here, $T_c$ is 
the critical temperature for the BCS transition and $k_BT_F=\mu_g$ is 
the Fermi temperature for the $\ket{g}$ atoms. For the present parameters,
 we have $k_BT_c\approx 2.8\omega$ giving 
$T_c^2/T_F\sim0.25$ in qualitative agreement with the results depicted in 
Fig.\ (\ref{mug315eempty}). We have performed numerical calculations for several values 
$g_{gg'}$ and $\mu_g$, and we find the general behavior as described in the 
present example. In all the tested cases, the current peak for the superfluid phase 
is shifted away to negative values of $\Delta$ and the shape of the peak is 
asymmetric as opposed to the simple Lorentzian shape for the normal phase. 
The shift of the peak is of the order ${\mathcal{O}}(T_c^2/T_F)$.

\begin{figure}
\centering
\epsfig{file=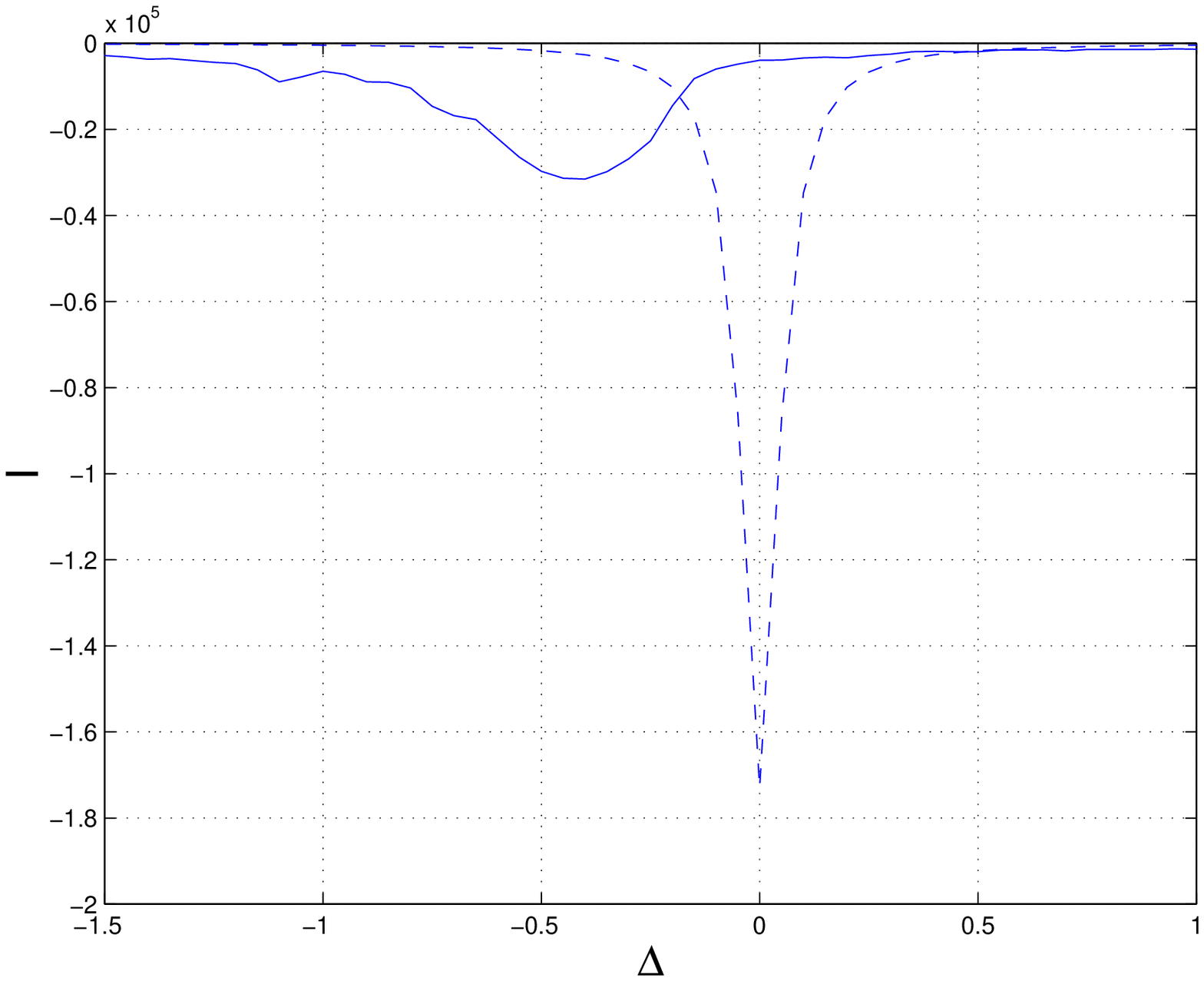,height=0.4\textheight,width=\textwidth}
\caption{The current $I=-\langle\dot{N}_e\rangle$. The solid/dashed lines are for the $\ket{g}$
and $\ket{g'}$ atoms in the superfluid/normal phase for $\mu_g=31.5$, $\mu_e=21.5$ 
(trap units), and  $g_{eg}+g_{eg'}=g_{gg'}$.}

\label{mug315mue215}
\end{figure}

In order to enhance the effect of the pairing on $I(\Delta)$ even
further, one could initially also trap some $\ket{e}$ atoms.  As long
as $\mu_g-\mu_e\gg\Delta_G$, the $\ket{e}$ atoms will not Cooper pair
with the $\ket{g}$ or $\ket{g'}$ atoms even though $g_{eg}$ or
$g_{eg'}<0$~\cite{Stoof}. By having the lower QP states for the
$\ket{e}$ atoms filled, there will only be transitions between the QP
states around the Fermi energy. Since these states are influenced the
strongest by the pairing, the effect of the superfluidity on
$I(\Delta)$ will be even stronger than for the parameters relevant for
Fig.\ \ref{mug315eempty}.  This is illustrated by Fig.\ 
\ref{mug315mue215}, where we plot $I(\Delta)$ for the same parameters
as above apart from that now $\mu_e=21.5\hbar\omega$. We see that both
the asymmetry and the shift in the peak in the superfluid phase as
compared to the normal phase, is more pronounced than in Fig.\ 
(\ref{mug315eempty}). This is simply because the transitions deep
below the Fermi level which are essentially inert to the effects of
superfluidity, are blocked by filling up the levels for the $\ket{e}$
atoms up to the energy $E=21.5\hbar\omega$. However, it might be
somewhat more difficult to achieve this situation experimentally, as
it requires the initial trapping of 3 (instead of 2) hyperfine states
with a rather good control of the populations in each state.

\begin{figure}
\centering
\epsfig{file=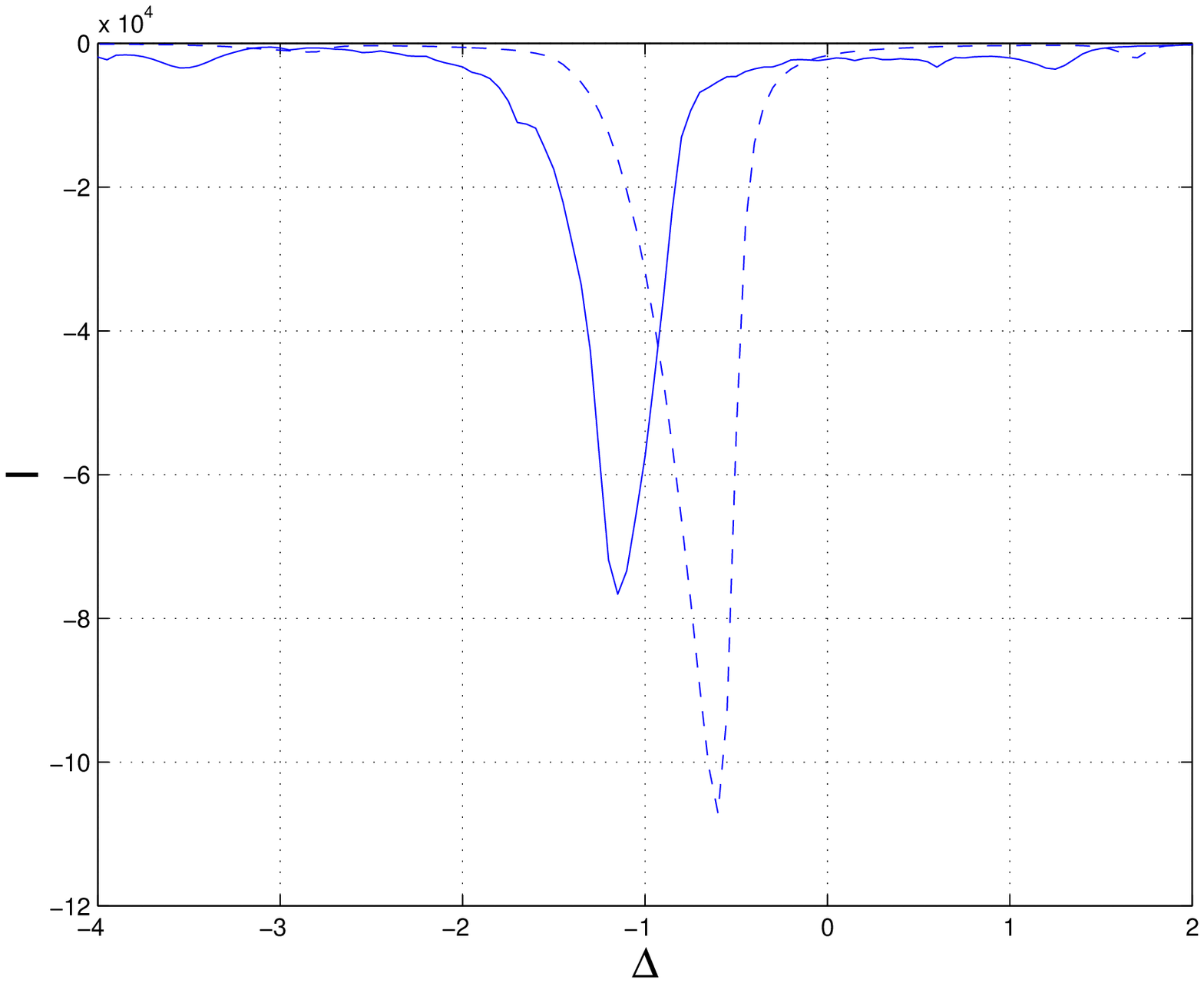,height=0.4\textheight,width=\textwidth}
\caption{The current $I=-\langle\dot{N}_e\rangle$. The solid/dashed lines are for the $\ket{g}$
and $\ket{g'}$ atoms in the superfluid/normal phase for $\mu_g=31.5$, $\mu_e=0$ 
(trap units), and $g_{eg}+g_{eg'}=0.9\times g_{gg'}$.}

\label{mug315eempty09}
\end{figure}

One should note that it is important that the Hartree field seen by the $\ket{g}$ and $\ket{e}$
atoms is approximately the same.  Otherwise, the wavefunctions and the spectra
for the $\ket{g}$ and $\ket{e}$ atoms will be different even when the $\ket{g}$ and $\ket{g'}$ atoms are 
in the normal phase. The overlap integrals will then  not give simple selection rules and there will be 
a contribution from many Lorentzians centered in general away from $\Delta=0$. Consequently,
$I(\Delta)$ will not be 
given by the simple formula in Eq.\ (\ref{Lorentz}). This is illustrated in 
Fig.\ \ref{mug315eempty09},
 where we plot the current $I(\Delta)$ for the same parameters as given above (with no $\ket{e}$ atoms 
initially), apart from we now have $g_{eg}+g_{eg'}=0.9\times g_{gg'}$. As expected,
 the current profile is shifted away from $\Delta=0$ and is asymmetric, even when 
the $\ket{g}$ and $\ket{g'}$ atoms are 
in the normal phase. The shift to negative frequencies is easy to understand: The attractive mean (Hartree) field 
seen by the $\ket{e}$ atoms is smaller than the attractive field seen by the $\ket{g}$ atoms since 
 $g_{eg}+g_{eg'}=0.9\times g_{gg'}$. Therefore, the trap states for the $\ket{e}$ atoms in general have a slightly 
higher energy than for the $\ket{g}$ atoms, and the normal phase current is shifted to negative $\Delta$.
 Figure \ref{mug315eempty09} demonstrates that the pairing field still causes a general 
shift of $I(\Delta)$ to negative frequencies and introduces further asymmetry since the pairing energy 
still needs to be broken to 
generate a current from the superfluid phase. This effect is  readily visible since the Hartree fields seen 
by the $\ket{g}$ and $\ket{e}$ atoms are approximately the same for the parameters chosen. However, if 
the difference in the Hartree fields becomes too large, the spread in the signal is determined by this 
difference and any  additional effects coming from the pairing field are correspondingly obscured.
 In  general, to be able to detect the presence of superfluidity using the scheme described in this 
section, one should have $\Delta W\ll\Delta_G$ where 
 $\Delta W=|g_{gg'}-g_{eg}-g_{eg'}|\rho$ denotes the difference in the Hartree fields with  $\rho$ being the 
average density of the $\ket{g}$ atoms, and $\Delta_G$ is the average gap. 

We conclude that if  $g_{eg}+g_{eg'}=g_{gg'}$ to a very good approximation so that 
 the difference in the Hartree fields seen by the $\ket{g}$ and the $\ket{e}$ is negligible,
the effect of superfluidity on the current $I(\Delta)$ should be straightforward to observe.
 The current in the normal phase is a simple Lorentzian centered 
around $\Delta=0$, whereas in the superfluid phase it is asymmetric and 
shifted away from $\Delta=0$. Furthermore, the shift in the center of the peak 
provides an estimate of $T_c$ if $T_F$ is known. Both the asymmetry and the shift away 
from $\Delta=0$ should be easily observable indications of the presence of superfluidity.
The effect is further enhanced if one initially traps $\ket{e}$ atoms keeping 
$\mu_g-\mu_e\gg\Delta_G$. In general, the scheme described in this section 
requires that the difference in the Hartree fields 
seen by the $\ket{g}$ and the $\ket{e}$ is smaller than the average pairing field in order to obtain a
 visible effect of the superfluidity on the current. 

\subsection{Equal chemical potentials}\label{mue=mug}

\begin{figure}
\centering
\epsfig{file=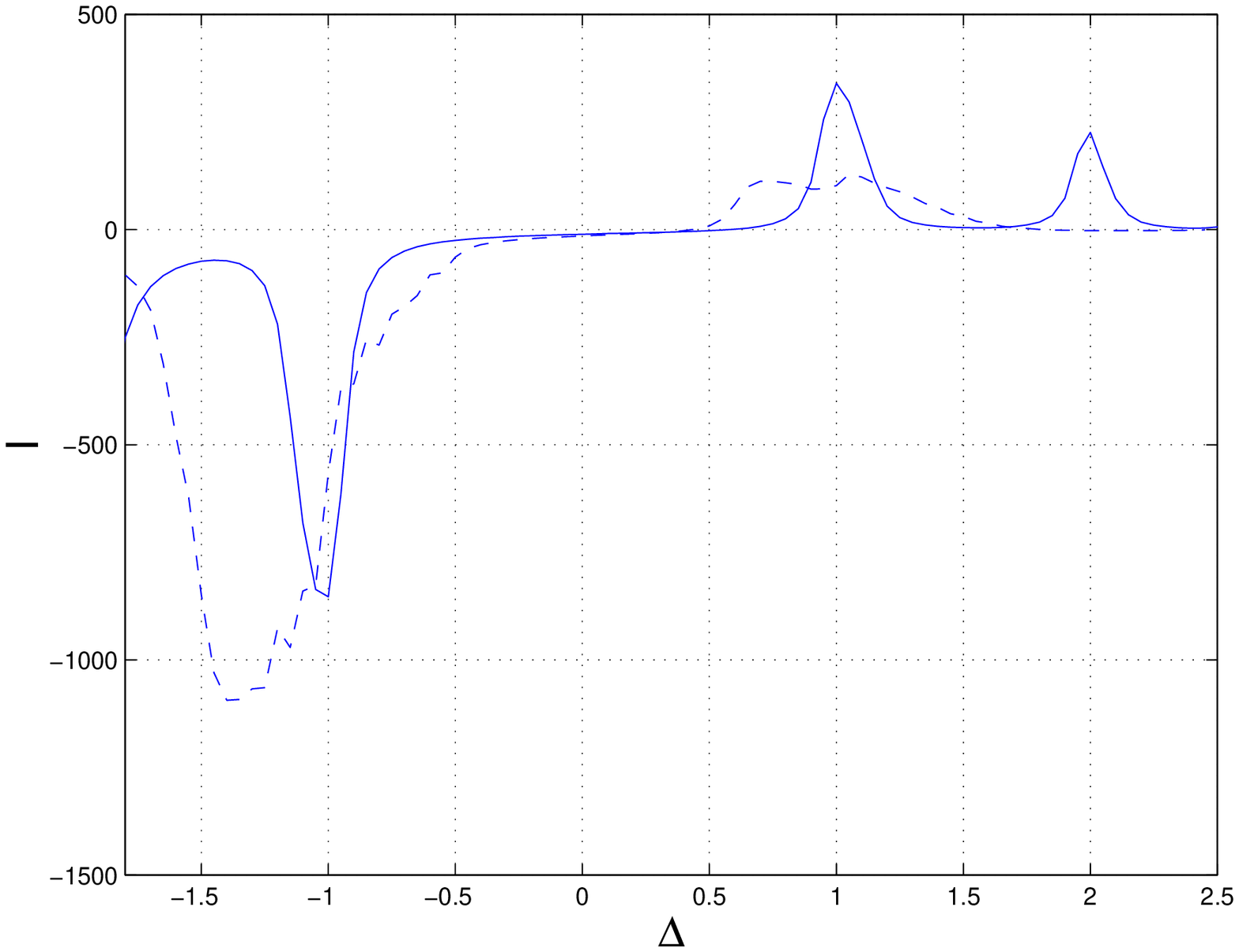,height=0.45\textheight,width=\textwidth}
\caption{The current $I=-\langle\dot{N}_e\rangle$. The solid/dashed lines are for the $\ket{g}$
and $\ket{g'}$ atoms in the superfluid/normal phase. Here $\mu_g=\mu_e=31.5$ (trap units), 
$g_{eg}=g_{eg'}=0$.}
\label{mugmue315gN0}
\end{figure}

We now consider the case of $\mu_e\simeq\mu_g$, where there is initially many atoms trapped in the 
state $\ket{e}$. Hence, to avoid the otherwise interesting possibility 
of the $\ket{e}$ atoms participating in the pairing, we assume that $g_{eg}=g_{eg'}=0$. 
In Fig.\  \ref{mugmue315gN0}, we plot the current $I(\Delta)$ for $g_{gg'}=-l_{ho}^3\omega$,
 $\mu_e=\mu_g=31.5\omega$,
 T=0, and $g_{eg}=g_{eg'}=0$. We have taken $\Gamma=0.1\omega$. As can be seen, there 
are several peaks in $I(\Delta)$ both when  $\ket{g}$ and $\ket{g'}$ are in the normal phase 
and when they are in the superfluid phase. In both cases, the peaks simply correspond to 
individual QP energy bands overlapping.

\begin{figure}
\centering
\epsfig{file=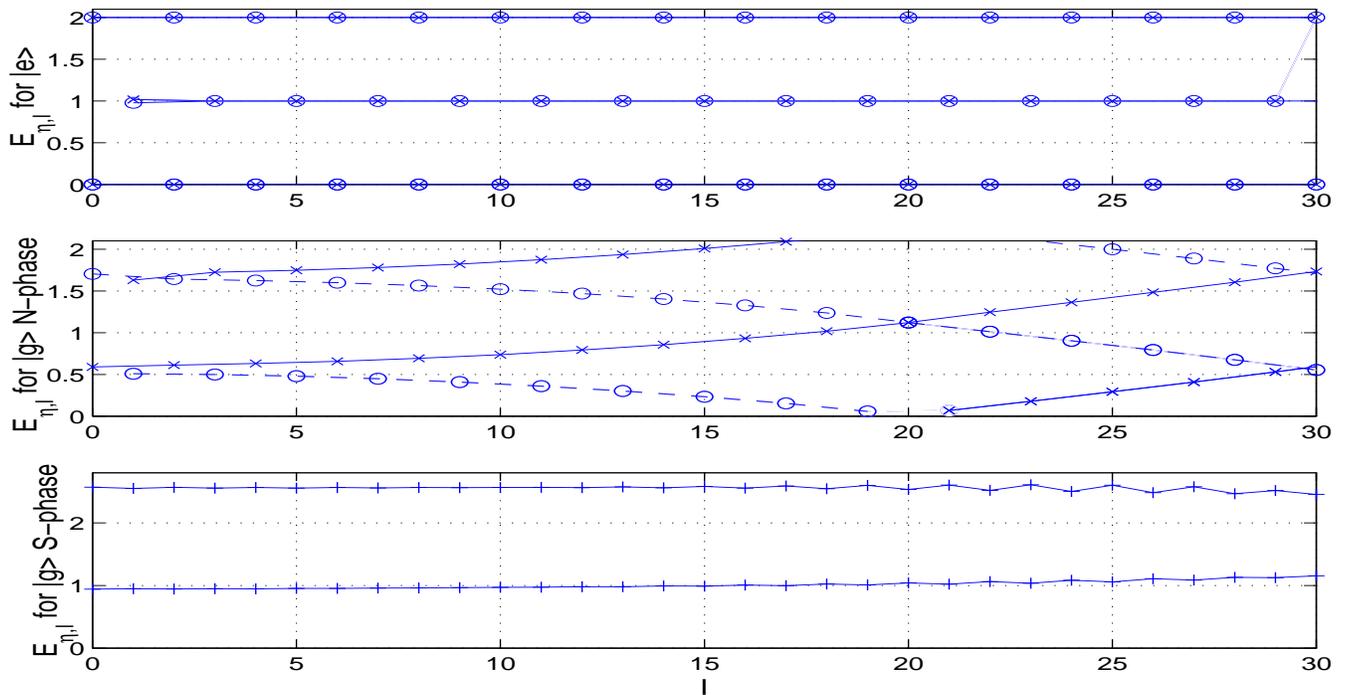,height=0.4\textheight,width=\textwidth}
\caption{The lowest QP energies $E_{\eta,l}$ as a function of the angular momentum $l$,
parameters are the same as in Fig.\ 4.}
\label{QPEnergiesall}
\end{figure}

To see this, we plot in Fig.\ \ref{QPEnergiesall} the corresponding lowest QP
energies  for the $\ket{e}$, $\ket{g}$, and $\ket{g'}$ atoms in both phases as a function 
of their angular momentum $l$. As we are 
in the Bogoliubov picture, all QP energies are positive and measured relative to the 
chemical potential. In the normal phase, negative QP energies $\xi_n=\epsilon_n-\mu<0$ 
are represented as positive QP energies $E_n=|\xi_n|$ with hole character. In 
Fig.\ (\ref{QPEnergiesall}), we label a hole-state by $\circ$ whereas a 
particle state is indicated by a $\times$. In the superfluid phase, the QP's are 
in general a superposition of a hole and a particle which we label by $+$. 

We see that when the Hartree field is attractive, a normal phase
energy band with a downward curvature in Fig.\ \ref{QPEnergiesall}
is a hole band whereas a normal phase energy band with an upward
curvature is a particle band. The reason is that particle states with
lower angular momentum $l$ has a lower energy for an attractive
interaction ($g<0$) than states with a higher $l$ since the
wavefunction overlap with the Hartree field decreases with increasing
$l$~\cite{BruunBurnett}.

The QP bands for the $\ket{e}$ atoms are  flat as they are the simple 
unperturbed harmonic oscillator states with energies $E_n=|(n+3/2)\omega-\mu_e|$.
 The lowest $E=0$ band corresponds to the harmonic oscillator states at the chemical potential 
($n=30$) with angular momentum $l=0,2\ldots,30$.  The interpretation of the spectra is described
in detail in \cite{BruunBurnett,Bruun99}. There is a one to one correspondence between the
 peaks in Fig.\ \ref{mugmue315gN0} and the QP bands depicted in Fig.\ \ref{QPEnergiesall}.
 For instance, the broad peak
 centered around  $\Delta=\omega$ when the $\ket{g}$ and $\ket{g'}$ atoms are in the normal
 phase corresponds to transitions from the half-filled QP band at $E=0$ for the $\ket{e}$ 
atoms into the empty (particle) band with $0.5\lesssim E/\omega\lesssim1.8$ for $l=0,2\ldots 30$ 
for the $\ket{g}$ atoms in the normal phase. 

We note that the peaks for the superfluid phase are 
sharper than the peaks for the normal phase. This is because the lowest QP bands for the $\ket{g}$
atoms are almost degenerate as a function of $l$ in the superfluid phase as can be seen from 
Fig.\ \ref{QPEnergiesall}. These states are strongly influenced
by the pairing field, which ``pushes'' them away from the center of the trap. They are 
concentrated in the region between where the pairing field and the trapping potential are 
significant \cite{Baranov98,Bruun99}, and they thus do not feel the Hartree field. Therefore,
their dependence on the quantum number $l$ is much weaker than in the normal phase.

We have performed calculations for a number of different 
parameters and we have observed the general behavior of $I(\Delta)$ as described above.
We conclude that when we have $\mu_e\simeq\mu_g$ and $g_{eg}=g_{eg'}=0$, the presence of
superfluidity is somewhat harder to detect as compared to the situation described in 
sec.\ \ref{mue0}. This is because the Hartree field tends to obscure
 any additional effect coming for the pairing. 
The current $I(\Delta)$ has in general many peaks corresponding to the energetic overlap
between individual QP bands for the $\ket{e}$ and the $\ket{g}$ atoms both for
the normal and the superfluid phases. The effect of the superfluidity is to make the peaks 
 sharper  than in the normal phase since the pairing field tends restore the degeneracy in the QP 
energies with respect to the angular momentum. One could therefore possibly
detect the onset of superfluidity as a sharpening of the peaks. 

\section{Probing the center of the trap -- effectively homogeneous system}\label{Sphere}

We now assume that the laser intensity $\Omega({\mathbf{x}})$ is large in the
 center of the cloud and that it decreases quickly as a function of the
distance from the center of the trap. This situation can be experimentally
achieved by using a Raman transition scheme with two perpendicular laser beams 
crossing each other at the center of the cloud. 
The profile of each laser beam should be narrow on the length scale of the 
trapped cloud. Since the laser beams then effectively probe only atoms 
in the center of the cloud where the Hartree and pairing fields are 
approximately constant, we would expect the observed signal to be well 
described by the results for a homogeneous system as given by Eq.(\ref{peak}). From
section \ref{Homogeneous}, we conclude that, for a homogeneous system, 
the optimal way of detecting the presence 
of the pairing field is to have $\mu_g\simeq\mu_e$ (see Eq.(\ref{curr2})). In this limit,
all low lying transitions far away from the Fermi energy and thus very little affected 
by the pairing, are Fermi blocked. The transitions contributing to $I(\Delta)$ are all
close to the Fermi level and hence strongly influenced by the presence of the pairing 
field. 
We will therefore concentrate on the case where the effective chemical potential 
in the center of the trap is the same for $\ket{g}$ and $\ket{e}$. As we will see, this 
case opens up the interesting possibility of directly measuring the size of the gap in the
 center of the cloud. 

\begin{figure}
\centering

\epsfig{file=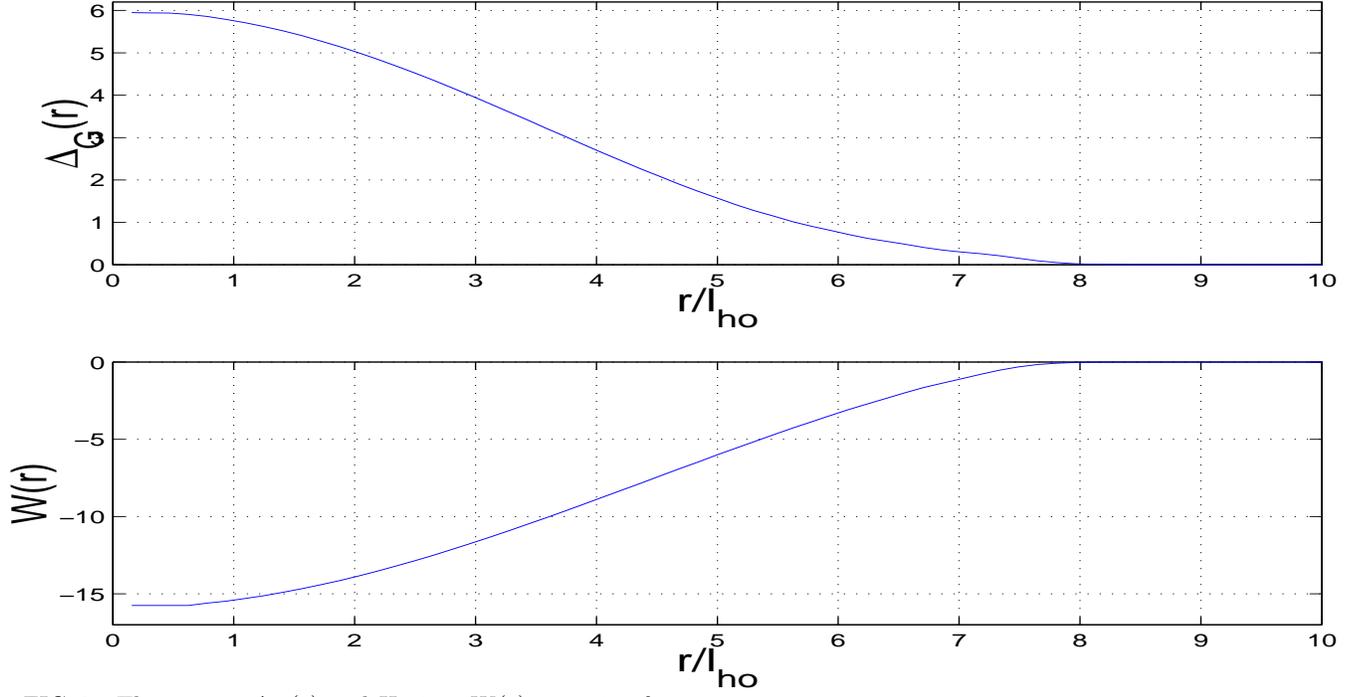,height=0.4\textheight,width=\textwidth}
\caption{The pairing $\Delta_G(r)$ and Hartree $W(r)$ in units of $\omega$, parameters
are the same as in Fig.\ 4.}
\label{DelHar}
\end{figure}

\begin{figure}
\centering
\epsfig{file=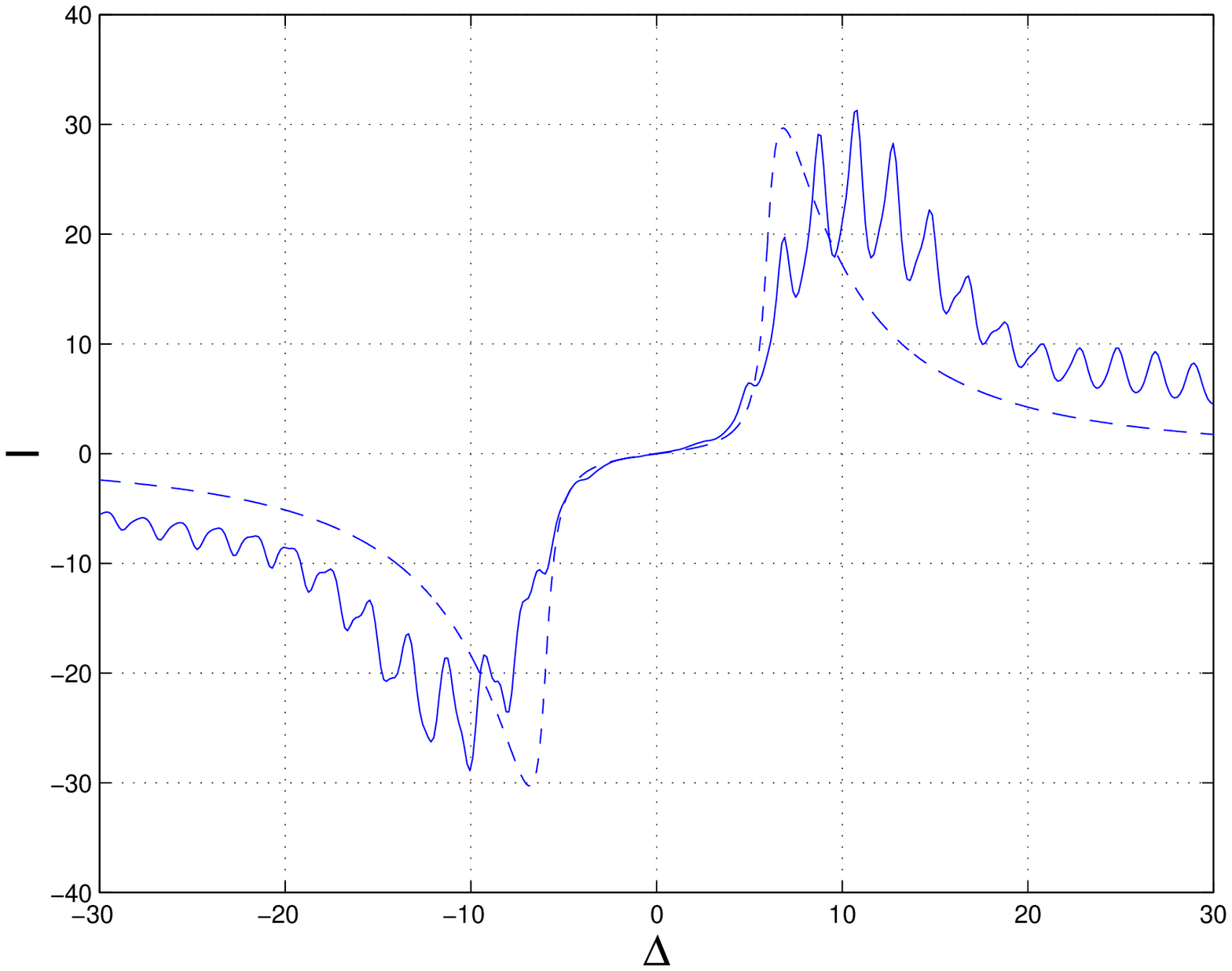,height=0.4\textheight,width=\textwidth}
\caption{The current $I=-\langle\dot{N}_e\rangle$ for $\mu_g=\mu_e=46.5$ (trap units)
and $g_{eg}=g_{eg'}=0$. The solid line is the exact numerical 
result whereas the dashed line is based on Eq.(\ref{curr2}), with $\Delta_G = 6$.}
\label{muemugTopHat}
\end{figure}

We use the same set of parameters as in sec.\ \ref{mue=mug} for the $\ket{g}$ and $\ket{g'}$ 
atoms. But now we assume that the
intensity profile for the beam can be well approximated by a sphere of constant intensity 
for $r\le r_0$ and  zero intensity for $r>r_0$ with $r$ denoting 
the distance to the center of the trap. That is, we take 
$\Omega({\mathbf{r}})=\Omega\Theta(r_0-r)$ in Eq.(\ref{final}) with $r_0=2l_{ho}$. 
>From Fig.\ \ref{DelHar},
we see that $\Delta_G(r)\simeq6\omega$ and $W(r)\simeq15\omega$ for $r\le2l_{ho}$. 
Thus, the effective local potential for the $\ket{g}$ atoms is $\mu_g\sim 46.5\omega$.
We therefore take $\mu_e=46.5\omega$ and $g_{eg}=g_{eg'}=0$ in order to have the same
effective local 
chemical potential for $\ket{g}$ and $\ket{e}$ in the center of the trap. 
The result is shown in Fig.\ \ref{muemugTopHat}.
Since the system is approximately homogeneous for $r\le2l_{ho}$, one 
could expect the current profile to be well described by Eq.(\ref{curr2}), with
$\mu_e=\mu_g=46.5\omega$ and $\Delta_G=6\omega$. 
We therefore also plot the result predicted by Eq.(\ref{curr2}). Here we take 
$\mu_e=\mu_g=46.5\omega$ and $\Delta_G=6\omega$ and we have normalized $I(\Delta)$ to a 
volume of $V=4\pi r_0^3/3$. We have taken a rather large value 
of $\Gamma=1\omega$ such that the discrete nature of the trap spectrum is 
washed out. Note that a finite imaginary part $\Gamma$ to the QP energies 
corresponds to convoluting Eq.(\ref{curr2}) with a Lorentzian of width $\Gamma$. 

As can be seen, there is a good agreement between the exact numerical
result and the prediction based on Eq.(\ref{curr2}). Especially, the current is zero for 
$-6\omega\lesssim\Delta\lesssim6\omega$ as predicted by Eq.(\ref{curr2}), since 
one either needs to break a Cooper pair with pairing energy $\sim \Delta_G(r=0)$ to 
produce a current into $\ket{e}$ ($\Rightarrow\Delta\lesssim-6\omega$)   
or one has to create a QP with energy minimum $\sim \Delta_G(r=0)$
($\Rightarrow\Delta\gtrsim6\omega$) to generate a current into $\ket{g}$.
The peaks in the numerical result reflect, of course, the discreteness of the  trap levels.
These peaks are quite large, as there is only $\approx500$ 
particles trapped in the region $r\le2l_{ho}$ for the parameters given above. Clearly, these 
peaks cannot be reproduced by the homogeneous treatment given by Eq.(\ref{curr2}) which however 
reproduces the general shape of the current profile well. If we had chosen a larger
system the individual peaks would be more numerous and smaller on the scale of 
the gap $\Delta_G(r=0)$ and the agreement between the homogeneous approximation and 
the exact result would probably be even better. 

We conclude that by concentrating the beam intensity to the center of the cloud where the 
gas can be to a good approximation regarded as homogeneous, the current profile $I(\Delta)$ is 
well described by the results presented in section \ref{Homogeneous}. An important result is 
that  by adjusting the parameters such that the local chemical potentials are the same 
 $[\mu_g-W_g(r=0)\simeq\mu_e-W_e(r=0)]$, one should be able to measure directly the size of the
 gap in the 
center of the cloud. It is simply given by the threshold detuning energy below which the 
observed current should be zero: for $|\Delta|\lesssim \Delta_G$
the current $I(\Delta)\simeq0$.

\section{Conclusions}\label{Conclusions}

The observation of the predicted BCS-state in gases of trapped atomic
Fermions poses a double challenge: the order parameter is small thus a
sensitive probe has to be found, furthermore, the trapping potential
leads to appearance of in-gap low-energy excitations which may make it
difficult to resolve the gap energy. In this paper we have presented a
method which is based on the transfer of atomic population over a
superconductor - normal state interface. This interface is effectively
created by using a laser to couple internal states with large and
small scattering lengths. The population transfer requires to break a
Cooper pair and the extra energy for this is provided by the laser
detuning. The change in the atomic population as a function of the
laser detuning thus gives information about the gap energy. We have
derived the current of population both in the case of a trapped gas
and a homogeneous system, and investigated the feasibility of the
method in different physical regimes.

We found that, in the case of a constant laser profile, the clearest
signatures of the BCS-state are observed when it is assumed that there
are initially no atoms in the normal state. Furthermore, the
scattering length between the normal state atoms and the Cooper paired
ones is assumed to be about the half of the scattering length between
the two Cooper paired ones -- this causes all the atoms to see the same
Hartree field. In this physical situation the effect of the BCS-state
is particularly simple and clear: the maximum in the current of
population as a function of the detuning is shifted and the peak
becomes asymmetric. Although this would probably be the optimal
choice, other initial conditions and probe parameters lead to clear
signatures of the BCS-state as well.

To avoid the problems arising from the non-homogeneous trapping
potential we propose to probe only the middle of the trap. The order
parameter is effectively homogeneous in the middle and the
wave-functions of the in-gap excitations are located away from the
center. In practice this kind of probing can be done by using two
orthogonal Raman beams which intersect only in the middle of the trap.
We have shown that indeed this leads to a result very similar to the
one in the homogeneous: the maximum of the current is shifted exactly 
by the amount of the gap energy. This allows a direct measurement of 
the gap energy.

{\it Acknowledgements} 
PT and MR acknowledge the support by the Academy of Finland
(Projects No.\ 48845, No.\ 47140 and No.\ 44897 Finnish Centre of 
Excellence Programme 2000-2005). Work at the
University of Innsbruck is supported by the Austrian Science Foundation and
EU TMR networks. P.T.
started this work as a Marie Curie Fellow of the EU in the University of Innsbruck.

\end{multicols}


\begin{thebibliography}{cc}
\bibitem{Jin99}
B.\ DeMarco and D.S.\ Jin, Science {\bf 285}, 1703 (1999).
\bibitem{Mewes00}
M.-O.\ Mewes, G.\ Ferrari, F.\ Schreck, A.\ Sinatra, and C.\ Salomon,
Phys.\ Rev.\ A \textbf{61}, 011403(R) (2000).
\bibitem{Thomas00}
K.M.\ O'hara, M.E.\ Gehm, S.R.\ Granade, S.\ Bali, and J.E.\ Thomas,
Phys.\ Rev.\ Lett.\ {\bf 85}, 2092 (2000).
\bibitem{BruunBurnett}
G.\ M.\ Bruun and K.\ Burnett, Phys.\ Rev.\ A \textbf{58}, 2427 (1998).
\bibitem{QHE}
Tin-Lun\ Ho, cond-mat/0005508.
\bibitem{BruunClark}
G.\ M.\ Bruun and C.\ W.\ Clark, Phys.\ Rev.\ Lett.{\bf 83}, 5415 (1999).
\bibitem{Stringari}
L.\ Vichi and S.\ Stringari, Phys.\ Rev.\ A \textbf{60}, 4734 (1999). 
\bibitem{Csordas}
A.\ Csord\'as  and R.\ Graham, cond-mat/0007049.
\bibitem{Coll}
G.\ Ferrari, Phys.\ Rev.\ A {\bf 59}, R4125 (1999); L.\
Vichi, cond-mat/0006305.
\bibitem{Janne99}
J.\ Ruostekoski and J.\ Javanainen, Phys.\ Rev.\ Lett.\ {\bf 82},
4741 (1999).
\bibitem{Zoller}
T.\ Busch, J.\ R.\ Anglin, J.\ I.\ Cirac and P.\ Zoller, 
Europhys.\ Lett.\ {\bf 44}, 1 (1998).
\bibitem{Stoof}
H.T.C.\ Stoof, M.\ Houbiers, C.A.\ Sackett, and R.G.\ Hulet, 
Phys.\ Rev.\ Lett.\ {\bf 76}, 10 (1996).
\bibitem{You99}
L.\ You and M.\ Marinescu, Phys.\ Rev.\ A {\bf 60}, 2324 (1999).
\bibitem{Bose}
For recent reviews, see E.A.\ Cornell, J.R.\ Ensher, and C.E.\ Wieman,
cond-mat/9903109; W.\ Ketterle, D.S.\ Durfee, and D.M.\ Stamper-Kurn,
cond-mat/9904034.
\bibitem{Weig99}
F.\ Weig and W.\ Zwerger, 
Europhys.\ Lett.\ {\bf 49}, 282 (2000).
\bibitem{Zhang99}
W.\ Zhang, C.A.\ Sackett, and R.G.\ Hulet,
Phys.\ Rev.\ A {\bf 60}, 504 (1999).
\bibitem{Ruostekoski99}
J.\ Ruostekoski, Phys.\ Rev.\ A {\bf 60}, R1775 (1999).
\bibitem{Baranov98}
M.A.\ Baranov,
JETP Lett.\ {\bf 70}, 396 (1999).
\bibitem{Baranov99}
M.A.\ Baranov and D.S.\ Petrov, 
cond-mat/9901108. 
\bibitem{Bruun99}
G.M.\ Bruun and C.W.\ Clark, 
cond-mat/990392 (to be published in J.\ Phys.\ B, Special issue on Coherent 
matter waves, November 2000).
\bibitem{Zambelli00}
F.\ Zambelli and S.\ Stringari, cond-mat/0004325.
\bibitem{Minguzzi00}
A.\ Minguzzi and M.P.\ Tosi, cond-mat/0005098.
\bibitem{Petrosyan99}
K.G. Petrosyan, JETP Letters {\bf 70}, 11 (1999).
\bibitem{PTPZ}
P.\ T\"orm\"a and P.\ Zoller, Phys.\ Rev.\ Lett.\ {\bf 85}, 487 (2000).
\bibitem{Mahan}
G.D.\ Mahan, {\it Many-Particle Physics} (Plenum Press, New York,
1990).
\bibitem{Bohn00}
J.L.\ Bohn, Phys.\ Rev.\ A {\bf 61}, 053409 (2000).
\bibitem{Williams99}
J.\ Williams, R.\ Walser, J.\ Cooper, E.\ Cornell, and M.\ Holland,
Phys.\ Rev.\ A {\bf 59}, R31 (1999).
\bibitem{deGennes}
P.G.\ de Gennes, {\it Superconductivity of Metals and Alloys},
(Addison-Wesley Publishing Company, 1989).
\bibitem{Georg} G.\ Bruun, Y.\ Castin, R. Dum, and K.\ Burnett, 
Eur.\ Phys.\ D \textbf{7}, 433 (1999).

\bibitem{Stoof97}
M.\ Houbiers, R.\ Ferweda, H.T.C.\ Stoof, W.I.\ McAlexander, C.A.\ Sackett,
and R.G.\ Hulet,
Phys.\ Rev.\ A {\bf 56}, 4864 (1997).
\bibitem{Abraham} E.\ R.\ I.\ Abraham, W.\ I.\ Alexander, 
J.\ M.\ Gerton, R.\ G.\ Hulet, R.\ Cot\'{e},
A. Dalgarno, Phys.\ Rev.\ A \textbf{55}, R3299 (1997).

\end{thebibliography}
\end{document}